\documentclass[usenat]{aa}
\usepackage{txfonts}
\usepackage{natbib}
\bibpunct{(}{)}{;}{a}{}{,}
\usepackage{longtable}
\usepackage[dvips]{graphicx}
\newcommand{\chandra}{\textit{Chandra}}
\newcommand{\sn}{$S_{\rm N}$}
\newcommand{\fxlf}{$f_{\rm XLF}$}

\begin{document}

\title{Dependence of the LMXB population on stellar age}
\author{Zhongli Zhang \inst{1} \and Marat Gilfanov \inst{1,2} \and {\'A}kos Bogd{\'a}n \inst{3} }
\institute{Max-Planck Institut f\"{u}r Astrophysik, Karl-Schwarzschild-Stra\ss e 1, D-85741 Garching, Germany 
\and Space Research Institute, Russian Academy of Sciences, Profsuyuznaya 84/32, 117997 Moscow, Russia 
\and Smithsonian Astrophysical Observatory, 60 Garden Street, Cambridge, MA 02138, USA 
\\
email:[zzhang;gilfanov]@mpa-garching.mpg.de}
\titlerunning{Stellar age dependence of LMXB population in early-type galaxies}
\date{Received ... / Accepted ...}

\abstract{} 
{We investigate the dependence of the low-mass X-ray binary (LMXB) population in early-type galaxies on 
stellar age.} 
{We selected 20 massive nearby early-type galaxies from the \chandra\ archive 
occupying a relatively narrow range of masses and spanning a broad range of ages, 
from 1.6 Gyr to more than 10 Gyrs, with the median value of 6 Gyrs. With the 
$\sim2000$ X-ray point sources detected in total, we correlated the specific 
number of LMXBs in each galaxy with its stellar age and globular cluster (GC) content.} 
{We found a correlation between the LMXB population and stellar age: older 
galaxies tend to possess about $\approx 50\%$ more LMXBs (per unit stellar mass) 
than the younger ones. The interpretation of this dependence is complicated by 
large scatter and a rather strong correlation between stellar age and 
GC content of galaxies in our sample. We present evidence suggesting that 
the more important factor may be the evolution of the LMXB population 
with time. Its effect is further amplified by the larger GC content of older 
galaxies and correspondingly, the larger numbers of dynamically formed binaries in them. 
We also found clear evolution of the X-ray luminosity function (XLF) with age, that younger 
galaxies have more bright sources and fewer faint sources per unit stellar mass. 
The XLF of LMXBs in younger galaxies appears to extend significantly 
beyond $10^{39}$ erg/s. Such bright sources seem to be less frequent in older 
galaxies. We found that 6 out of $\approx 12$ (ultra-) luminous sources are located 
in GCs.}
{}

\maketitle


\section{Introduction}
\label{sec:introduction}

LMXBs are accreting systems consisting of a
low-mass star ($\lesssim$1$M_\odot$) transferring mass onto a neutron
star or black hole through Roche-lobe overflow. Extensive studies of
nearby galaxies with \chandra\ have confirmed the long-suspected fact that
their contribution to X-ray emission in early-type galaxies is
substantial \citep[e.g.,][]{Irwin2003,Kim2004}. Their collective
luminosity was found to closely follow the near-infrared light as well as
the scaling relation of the LMXB population with the stellar mass 
obtained \citep{Gilfanov2004}. However, a moderate scatter exists
in these relations, suggesting that the specific number (per unit stellar mass) 
of X-ray binaries is not universally constant among galaxies and that 
secondary correlations may play a role. 

Obviously, one of the main candidates for the second-order correlation
is the LMXBs with the age of the stellar population. Unlike
high-mass X-ray binaries, LMXBs are found both in young and old
galaxies. Given that the characteristic time scale for the stellar
evolution of the donor star and for the orbital evolution of the
binary are both in the~Gyrs range, it is not surprising that younger
and older galaxies differ in their LMXB content. For example, 
\citet{Kim2010} recently reported evidence that younger galaxies
may have a higher fraction of bright sources than older ones. On
the theoretical side, the population synthesis calculations by
\citet{Fragos2008} predict that the formation rate of LMXBs 
steadily decreases with time after 1 Gyr from the star formation event. This 
conclusion seems to have been supported by observations: based on the 
analysis of galaxies detected in the extended \chandra\ Deep Field South, \citet{Lehmer2007}
found that for optically faint early-type galaxies (where LMXBs
dominate the X-ray emission), $L_{\rm X}/L_{\rm B}$ increases moderately 
with redshift over $z\sim0.0-0.5$ range. Furthermore, \citet{Fragos2012} 
calculated  the evolution of the specific LMXB luminosity through cosmic time. Their results suggest that the specific  luminosity density of LMXBs peaks at $z\sim2.5$ and declines towards redshift $z=0$. However, there are  observational facts that
appear to challenge these conclusions. In a S0 galaxy NGC 5102, whose 
stellar population is younger than 1 Gyr, \citet{Kraft2005} found only 
two sources brighter than $10^{37}$ erg/s, which is three times less than 
the predicted number of six LMXBs. \citet{Bogdan2010} reported similar results 
for two young elliptical galaxies, NGC 3377 and NGC 3585. Admittedly, 
both studies suffered from relatively low statistical significance of 
the results and therefore cannot be considered as a final argument. 
Similarly, the result of \citet{Lehmer2007} was not based on a direct
determination of the age of the stellar environment (which was rather
inferred from the redshift) and could have been contaminated by other
effects (e.g.,\ the rate of galaxy mergers). 

It is obvious that more observational effort is needed in order to
clarify this issue. The progress in this direction is hampered by
the difficulty in reliably determining the age of stellar
populations. In addition, a significant fraction of LMXBs in elliptical
galaxies reside in GCs that are dynamically formed in
two-body stellar interactions, rather than having a primordial
origin. In order to investigate the age effects on the primordial
population of LMXBs, GC sources need to be identified
and excluded completely from the analysis. To this end,  clean 
and reliable lists of GCs are needed, which are not available 
for the large number of galaxies required for a statistically meaningful study.

\begin{table*}
\begin{center}
\caption{The galaxy sample}
\label{tab:sample}
\begin{tabular}{rlrcrcccccc}
\hline
\hline
Galaxy& Type & Distance & $N_{\rm H}$           & $L_{\rm K}$                 & $M_{*}/L_{\rm K}$             & $r_{\rm e}$ & D25              & $M_{\rm V}$ & $N_{\rm GC}$ & \sn  \\
      &      & (Mpc)    & ($10^{20}$ cm$^{-2}$) &($10^{10}L_{\rm K,\odot}$) & ($M_{\odot}/L_{\rm K,\odot}$) & (arcmin)    & ($2a,2b,\theta$) & (mag)       &              &        \\ 
(1)   & (2)  &  (3)     &  (4)                  &  (5)                        &  (6)                          & (7)         & (8)              &  (9)        & (10)         & (11) \\ 
\hline 
\\
N720 & E5   &  27.7  &  1.54  & 21.50 & 0.86 & 1.20 &  $4.7\arcmin,2.4\arcmin,140\degr$    & -22.04 & $660\pm190^a$   & $1.01\pm0.29$    \\
N821 & E6   &  24.1  &  6.39  & 9.12  & 0.82 & 1.66 &  $2.6\arcmin,1.6\arcmin,25\degr$     & -21.12 & $320\pm45^b$    & $1.14\pm0.16$    \\
N1052 & E4   &  19.4  &  3.07  & 8.94  & 0.80 & 1.12 &  $3.0\arcmin,2.1\arcmin,120\degr$    & -21.00 & $400\pm45^c$    & $1.59\pm0.18$   \\
N1380 & SA0  &  17.6  &  1.31  & 12.57 & 0.81 & 1.32 &  $4.8\arcmin,2.3\arcmin,7\degr$      & -21.23 & $560\pm30^d$    & $1.81\pm0.10$   \\
N1404 & E1   &  21.0  &  1.36  & 18.73 & 0.85 & 0.79 &  $3.3\arcmin,3.0\arcmin,162.5\degr$  & -21.58 & $725\pm145^e$   & $1.69\pm0.34$   \\
N3115 & S0   &  9.7   &  4.32  & 9.43  & 0.83 & 1.07 &  $7.2\arcmin,2.5\arcmin,40\degr$     & -21.13 & $630\pm150^f$   & $2.22\pm0.53$   \\
N3379 & E1   &  10.6  &  2.75  & 7.92  & 0.83 & 1.17 &  $5.4\arcmin,4.8\arcmin,67.5\degr$   & -20.88 & $270\pm69^g$    & $1.20\pm0.31$  \\
N3585 & E6   &  20.0  &  5.58  & 18.92 & 0.77 & 1.20 &  $4.7\arcmin,2.6\arcmin,107\degr$    & -21.76 & --              & $0.50\pm0.15^h$ \\
N3923 & E4-5 &  22.9  &  6.21  & 29.90 & 0.82 & 1.66 &  $5.9\arcmin,3.9\arcmin,50\degr$     & -22.11 & $2494\pm286^i$  & $3.57\pm0.41$   \\
N4125 & E6   &  23.9  &  1.84  & 23.49 & 0.80 & 1.95 &  $5.8\arcmin,3.2\arcmin,82.5\degr$   & -22.13 & --              & $1.30\pm0.50^h$ \\
N4278 & E1-2 &  16.1  &  1.77  & 7.87  & 0.78 & 1.15 &  $4.1\arcmin,3.8\arcmin,27.5\degr$   & -20.96 & $1300\pm300^f$  & $5.35\pm1.23$   \\
N4365 & E3   &  20.4  &  1.62  & 20.86 & 0.85 & 1.66 &  $6.9\arcmin,5.0\arcmin,40\degr$     & -22.01 & $2511\pm1000^j$ & $3.95\pm1.57$   \\
N4374 & E1   &  18.4  &  2.60  & 24.94 & 0.83 & 1.70 &  $6.5\arcmin,5.6\arcmin,135\degr$    & -22.25 & $4301\pm1201^k$ & $5.39\pm1.50$   \\
N4382 & SA0  &  18.5  &  2.52  & 27.06 & 0.76 & 1.82 &  $7.1\arcmin,5.5\arcmin,12.5\degr$   & -22.23 & $1110\pm181^k$  & $1.43\pm0.23$   \\
N4472 & E2   &  16.3  &  1.66  & 41.88 & 0.85 & 3.47 &  $10.2\arcmin,8.3\arcmin,155\degr$   & -22.68 & $7813\pm830^k$  & $6.61\pm0.70$   \\
N4552 & E0-1 &  15.3  &  2.57  & 10.82 & 0.83 & 0.98 &  $5.1\arcmin,4.7\arcmin,150\degr$    & -21.29 & $984\pm198^k$   & $2.99\pm0.60$   \\
N4636 & E0-1 &  14.7  &  1.81  & 13.24 & 0.81 & 2.95 &  $6.0\arcmin,4.7\arcmin,150\degr$    & -21.33 & $4200\pm120^l$  & $12.38\pm0.35$  \\
N4649 & E2   &  16.8  &  2.20  & 32.44 & 0.85 & 2.29 &  $7.4\arcmin,6.0\arcmin,105\degr$    & -22.38 & $4745\pm1099^k$ & $5.32\pm1.23$   \\
N4697 & E6   &  11.7  &  2.12  & 8.82  & 0.77 & 2.40 &  $7.2\arcmin,4.7\arcmin,70\degr$     & -21.16 & $1100\pm400^m$  & $3.78\pm1.37$   \\
N5866 & SA0  &  15.3  &  1.46  & 9.47  & 0.72 & 1.35 &  $4.7\arcmin,1.9\arcmin,128\degr$    & -20.93 & $400\pm100^n$   & $1.69\pm0.42$   \\
\hline
\end{tabular}
\end{center}
 (1) -- Galaxy name. (2) -- Morphological type. (3) -- Distance derived 
 from the surface brightness fluctuation method
 \citep{Tonry2001}. (4) -- Galactic column density
 \citep{Dickey1990}. (5) -- Total $K_{\rm S}$-band luminosity
 calculated from the total apparent $K_{\rm S}$-band magnitude from the 
 2MASS Large Galaxy Atlas \citep{Jarrett2003} and the distance adopted in this
 paper. (6) -- $K_{\rm S}$-band mass-to-light ratios derived from
 \citet{Bell2001}, with $B-V$ colors from RC3 catalog
 \citep{Dev1991}. (7) -- Effective radius from $B$-band photometry in
 RC3 catalog. (8) -- D25 region of major diameter ($2a$), minor
 diameter ($2b$), and position angle ($\theta$) from RC3 catalog. For
 NGC 1404, NGC 3379, NGC 4125, NGC 4278, NGC 4382, and NGC 4552, the
 position angle is taken from $K_{\rm S}$-band image. (9) -- Absolute
 $V$-band magnitude calculated from $m_{\rm v}$ from RC3. (10) --
 Total number of GCs. References --
 $^a$\citet{Kissler1996}; $^b$\citet{Spitler2008};
 $^c$\citet{Forbes2001}; $^d$\citet{Kissler1997};
 $^e$\citet{Forbes1998}; $^f$\citet{Harris1991};
 $^g$\citet{Rhode2004}; $^i$\citet{Sikkema2006};
 $^j$\citet{Forbes1996}; $^k$\citet{Peng2008}; $^l$\citet{Dirsch2005};
 $^m$\citet{Dirsch1996}; $^n$\citet{Cantiello2007}. (11) -- 
 Globular cluster specific frequency calculated from their total number 
 and total absolute $V$-band magnitude of the host galaxy, except for the 
 two galaxies for which local values of  \sn\ are used, from $^h$\citet{Humphrey2009}.  
\end{table*}

By now, \chandra\ has observed a large number of galaxies with different
morphological types and ages. On the other hand, significant progress
has been achieved in the accuracy of age-determination techniques
and advanced spectroscopical methods have been applied to a large
number of galaxies. This motivated us to undertake a systematic study
of the dependence of properties of LMXB populations on stellar
age. Among such properties, we consider the specific (per unit stellar
mass) number and X-ray luminosity of LMXBs and their luminosity
distributions. In our analysis, we will take into account possible
contamination by the GC sources to the degree allowed 
by the available GC data.   

The paper is structured as follows: In Sect.~\ref{sec:sample} we
describe our selection criteria and the resulting sample. In
Sect.~\ref{sec:data} we describe the X-ray and near-infrared data
preparation and analysis. In Sect.~\ref{sec:xlf} we discuss average 
scaling relations for LMXBs and their average XLFs. 
Dependence of the LMXB numbers and luminosity distribution on stellar 
age is discussed in Sect. \ref{sec:dependence}. In Sect.~\ref{sec:ulx} 
we consider the origin of the luminous X-ray sources in early-type galaxies 
and their dependence on age. In Sect.~\ref{sec:discussion} we 
discuss caveats and implications of our results. Finally, our results are summarized 
in Sect.~\ref{sec:summary}.

\section{The sample}
\label{sec:sample}

Our goal was to build the largest possible sample covering the widest
possible range of stellar ages. The size of the sample, however, was
limited by the content of the \chandra\ archive and by the published
age determinations. Our selection criteria were the
following. Firstly we selected all early-type (E/S0) galaxies
available in the \chandra\ archive. We cross-correlated this list 
with publications on stellar age determinations, leaving only 
galaxies for which reliable age determinations are available (see
below). From the remaining galaxies, we selected only the ones located
within the distance of $\sim25$ Mpc, which ensures a source detection
sensitivity of better than $5\cdot10^{37}$ erg/s in less than
150 ksec of \chandra\ observation. Then we chose massive systems with
$L_{\rm K}>5\cdot10^{10}L_{\rm K,\odot}$ to guarantee the presence of
a statistically meaningful number of LMXBs ($\gtrsim20$) above the
\chandra\ sensitivity limit. Finally, we excluded galaxies
with ongoing or very recent star formation since the stellar content
in such galaxies is likely to be inhomogeneous.  

In total, we selected 20 galaxies with the integrated $K_{\rm S}$-band luminosity
in the relatively narrow range from $\sim8\cdot10^{10}$ to $4\cdot10^{11}L_{\rm
  K,\odot}$. The main properties of these galaxies are listed in
Table~\ref{tab:sample}. \chandra\ detection sensitivity ($L_{\rm lim}$),
which is defined as 60\% completeness level (Sect.~\ref{sec:icf})
of LMXBs in the study field (Sect.~\ref{sec:detection}), ranges from
$\sim4\cdot10^{36}$ to $10^{38}$ erg/s (Table~\ref{tab:obs}). 
This ensures that there is a statistically meaningful number of compact 
sources in each galaxy.

\subsection{The stellar age}
\label{sec:age}

The most accurate and widely used method of age determination of
elliptical galaxies is the spectroscopic estimator, which compares
observed strength of absorption lines of age-sensitive elements
with predictions from the simple stellar population (SSP) synthesis
models. A number of such measurements for different galaxy samples are
published in the literature, e.g., \citet{Trager2000,Kuntschner2001,Terlevich2002,
Caldwell2003,Thomas2005,Denicolo2005,Sanchez2006,Annibali2007,Gallagher2008}. 
It is known that contamination by gas emission is one of the most important 
factors affecting the accuracy of  age determination. 
Therefore, for galaxies with more than one measurement, we chose those  
correcting gas emission in a more rigorous way, and prioritized the 
age-determination studies in the following order: 1) \citet{Annibali2007},
2) \citet{Sanchez2006}, and 3) \citet{Terlevich2002}. These three papers contain 
ages for 18 galaxies in our sample. They are summarized in Table~\ref{tab:age}, 
where for each galaxy we also list the adopted age. Two galaxies (NGC 3923 and 
NGC 4125) are not covered by these measurements. We adopted their ages from 
\citet{Thomas2005} for NGC 3923 and from \citet{Schweizer1992} for NGC 4125.

The accuracy and limitations of age determination and its impact on our results are discussed
in Sect.~\ref{sec:caveats_age}.

\begin{table}
\begin{center}
\caption{Stellar age measurements.} 
\label{tab:age}
\renewcommand{\arraystretch}{0.8}
\begin{tabular}{rccccc}
\hline
\hline
Galaxy &  Age1  &  Age2  &  Age3   &  Adopted age  \\
       & (Gyr)  & (Gyr)  & (Gyr)   & (Gyr) \\  
\hline
\\
$^{\rm Y}$N720  & 	       & 	      &  3.4  & 3.4                 \\  
$^{\rm Y}$N821  & 	       & $5.2\pm1.5$  &  7.2  & $5.2\pm1.5$	    \\  
N1052           & $14.5\pm4.2$ & 	      &       & $14.5\pm4.2$	    \\
$^{\rm Y}$N1380 & $4.4\pm0.7$  & 	      &       & $4.4\pm0.7$	    \\
$^{\rm Y}$N1404 & 	       & 	      &  5.9  & 5.9		    \\  
N3115           & 	       & $8.4\pm1.1$  &       & $8.4\pm1.1$	    \\
N3379           & 	       & $8.2\pm1.1$  &  9.3  & $8.2\pm1.1$	    \\  
$^{\rm Y}$N3585 & 	       & 	      &  3.1  & 3.1		    \\  
$^{\rm Y}$N3923 & 	       & 	      &       & $3.3\pm0.8$       \\
$^{\rm Y}$N4125 & 	       & 	      &       & $5.0$             \\
N4278           & 	       & $12.5\pm1.2$ &  10.7 & $12.5\pm1.2$	    \\  
N4365           & 	       & $7.9\pm1.2$  &       & $7.9\pm1.2$	    \\
N4374           & $9.8\pm3.4$  & $11.3\pm1.3$ &  11.8 & $9.8\pm3.4$	    \\  
$^{\rm Y}$N4382 & 	       & 	      &  1.6  & 1.6		    \\  
N4472           & 	       & $9.6\pm1.2$  &  8.5  & $9.6\pm1.2$	    \\  
$^{\rm Y}$N4552 & $6.0\pm1.4$  & $12.4\pm1.2$ &  9.6  & $6.0\pm1.4$	    \\  
N4636           & $13.5\pm3.6$ & $10.3\pm1.3$ &       & $13.5\pm3.6$	    \\
N4649           & 	       & $16.9\pm2.3$ &  11.0 & $16.9\pm2.3$	    \\  
N4697           & $10.0\pm1.4$ & $5.9\pm1.2$  &  8.2  & $10.0\pm1.4$	    \\  
$^{\rm Y}$N5866 & 	       & 	      &  1.8  & 1.8		    \\  
\hline
\end{tabular}
\end{center}
Stellar ages are listed according to priority: 1 -- \citet{Annibali2007}, 2 -- \citet{Sanchez2006}, 
3 -- \citet{Terlevich2002}. Ages for NGC 3923 and NGC 4125 are from  \citet{Thomas2005} and  \citet{Schweizer1992} respectively.  
Galaxies marked by ``Y'' are classified as young galaxies in the 
Sect.~\ref{sec:xlf_age},  the remaining galaxies  are classified as old. 
\end{table}

\subsection{The GC content}

To characterize the GC content of a galaxy, we use
the GC specific frequency (\sn), which is conventionally
defined by the relation $S_{\rm N}=N_{\rm GC}10^{0.4(M_{\rm V}+15)}$
\citep{Harris1991}. As this parameter is sensitive to the assumed
distance to the galaxy, the sensitivity limit, and the completeness of
the optical data, we collected the most accurate measurements of the
total number of GCs ($N_{\rm GC}$) in our galaxies and then computed
\sn\ with the distances used in this paper. For two galaxies, NGC
3585 and NGC 4125, $N_{\rm GC}$ were not available, and we used
the local \sn\ from \citet{Humphrey2009} as an approximation of its
global value. All values of \sn\ are listed in Table~\ref{tab:sample}, 
and caveats are discussed in Sect.~\ref{sec:caveats_gc}.

\section{Data analysis}
\label{sec:data}

\subsection{Data preparation and source detection}
\label{sec:detection}

\chandra\ observations of our sample galaxies are listed in
Table~\ref{tab:obs}. We reduced the data following the standard CIAO threads
(CIAO version 4.2; CALDB version 4.2.1). We did not exclude time
intervals for background flares since the benefit of the increased
exposure time outweighs the increased background. The energy range was
limited to 0.5-8.0 keV. We made exposure maps in this energy range,
assuming the single power-law model with $\Gamma=1.7$ under the
galactic absorption for each galaxy. To detect point sources, we used
CIAO task \textit{wavdetect} with the parameters adopted from
\citet{Voss2006,Voss2007}. Thresholds were set to $10^{-6}$, yielding
on average one false detection per $8.2\arcmin\times8.2\arcmin$ area
($10^6$ ACIS pixels).

\begin{table}
\caption{\chandra\ observations.}
\label{tab:obs}
\begin{tabular}{rlrrr}
\hline
\hline
Galaxy &  Observation ID	                 & Exposure    & $L_{\rm min}$ &  $L_{\rm lim}$ \\
(1)    &   (2)	                                 & (3)         &   (4)         &  (5)	        \\ 
\hline 
\\
N720   &  492,7062,7372$^*$,8448,8449            & 138.8  & 3.6  & 9.5 \\
N821   &  4006,4408,5691,5692,6310,              & 212.9  & 1.3  & 2.8 \\
       &  6313$^*$,6314                          &        &      &     \\	
N1052  &  5910		                         & 59.2   & 3.1  & 6.3 \\
N1380  &  9526		                         & 41.6   & 3.9  & 6.1 \\
N1404  &  2942,4174$^*$,9798,9799                & 114.5  & 2.4  & 11.7\\
N3115  &  2040,11268,12095$^*$                   & 153.2  & 0.34 & 0.70\\
N3379  &  1587,7073$^*$--7076                    & 337.0  & 0.06 & 0.42\\
N3585  &  2078,9506$^*$	                         & 94.7   & 2.3  & 4.1 \\
N3923  &  1563,9507$^*$	                         & 102.1  & 2.6  & 6.3 \\
N4125  &  2071		                         & 64.2   & 3.0  & 8.9 \\
N4278  &  4741,7077--7081$^*$                    & 470.8  & 0.32 & 0.88\\
N4365  &  2015$^*$,5921--5924,7224               & 195.8  & 1.0  & 2.5 \\
N4374  &  803,5908$^*$,6131                      & 115.5  & 0.84 & 4.9 \\
N4382  &  2016		                         & 39.7   & 3.5  & 6.3 \\
N4472  &  321$^*$,322,11274                      & 89.6   & 0.58 & 5.6 \\
N4552  &  2072		                         & 54.4   & 1.3  & 4.5 \\
N4636  &  323,324,3926,4415$^*$                  & 209.8  & 0.13 & 3.9 \\
N4649  &  785,8182$^*$,8507                      & 108.0  & 2.1  & 6.8 \\
N4697  &  784,4727--4730$^*$                     & 193.0  & 0.41 & 0.83\\
N5866  &  2879		                         & 33.7   & 2.1  & 4.7 \\
\hline
\end{tabular}
 (1) -- Galaxy name. (2) -- \chandra\ observation IDs. (3) -- Total
 exposure time of \chandra\ observations. (4) and (5) -- The 0.5-8 keV
 luminosity of the faintest source detected, and the luminosity corresponding to the 60\% 
 completeness in the study field (computed assuming that the spatial distribution of LMXBs 
 follows the $K_{\rm S}$-band light). The luminosities are in the units of $10^{37}$ erg/s. 
\end{table}

We corrected the offsets of galaxies with multiple observations
following \citet{Voss2007}, by using the point sources detected in 
each observation within the $4\arcmin$ radius of the telescope axis. The
observations were then shifted using CIAO task \textit{reproject\_events} 
to match the coordinate system of the reference (marked with an
asterisk in Table~\ref{tab:obs}). The images were 
combined and re-analyzed. We performed \textit{wavdetect} again
on the combined images to finalize the point sources in each galaxy. To
avoid the source crowding problem and the bias of the incompleteness
of LMXBs in the galaxy center, we excluded the central $a=5\arcsec$
ellipse region (with the eccentricity and position angle following
D25). We define the region outside the central $5\arcsec$ inside D25
as the study field throughout this paper. The total number of point
sources detected in the study field is listed in Table~\ref{tab:stat}.

To estimate the source counts, we applied circular aperture centered on
the central coordinates (output of \textit{wavdetect}) of each source. We
defined the source region as including 85\% of the local point spread
function (PSF) value. The PSF file was extracted by CIAO task
\textit{mkpsf} from each image, then combined together for multiple
observations. The background region was defined as three times the
radius of the source region. We excluded the source regions from background 
regions with overlapping neighboring sources. The source net counts 
(with the majority in the source region and minority in the background 
region), and errors were then computed by the equations (1) and (2) in 
\citet{Voss2007}. To convert the absorbed source count rates into unabsorbed 
luminosities in 0.5-8 keV, we assumed a power-law spectrum ($\Gamma=1.7$) with
galactic absorption. We listed the faintest source detected in each
galaxy in Table~\ref{tab:obs} and the total number of point sources
above $L_{\rm lim}$ in Table~\ref{tab:stat}.

\subsection{Cosmic X-ray background sources}
\label{sec:cxb}

We estimated the cosmic X-ray background (CXB) sources (most of which are background 
active galactic nucleus (AGN)), using the full band (0.5-10 keV) log($N$)-log($S$) 
distribution of CXB sources from \citet{Geo2008} 
and converted the flux to the 0.5-8 keV band, assuming a power-law
spectrum with a photon index of 1.4. The total number of CXB sources
among all detected point sources and point sources above $L_{\rm
  lim}$ in the study field are listed in Table~\ref{tab:stat} (the
model was corrected by the incompleteness function of CXB sources
derived in Sect.~\ref{sec:icf}). In most galaxies, CXB sources
contribute less than 15\% of the total X-ray population, except for NGC
3379 and NGC 4382, where the contribution is somewhat higher (25-30\%), 
however, essential statistics sustain for the LMXB study. As well known, 
CXB source density is subject to field-to-field
variations due to the cosmic variance. These variations limit the
accuracy of the CXB level predictions based on the source counts in
selected extragalactic fields to $\sim 10-30\%$ of the predicted CXB value, 
depending on the solid angle. As the CXB contribution to the total number 
of sources is rather small, these uncertainties are relatively unimportant 
in most of the luminosity range. The situation changes  in the 
bright end of the XLF, where the cosmic variance 
becomes the major limiting factor in our analysis.

\begin{table}
\begin{center}
\caption{Statistics of point sources in the study field.} 
\label{tab:stat}
\begin{tabular}{rrrrrrrr}
\hline
\hline
Galaxy & $N^{\rm total}_{\rm X}$ & $N^{\rm total}_{\rm CXB}$ & $N^{L_{\rm lim}}_{\rm X}$ & $N^{L_{\rm lim}}_{\rm CXB}$ & $N^{L_{\rm lim}}_{\rm LMXB}$ & $L_{\rm K}$ & $M_{*}$   \\
 (1)   & (2) & (3) & (4)  & (5)   & (6)   & (7)  & (8)	 \\
\hline
\\
N720  & 79  & 5.9  & 60  & 4.8  & 60.8  & 19.01 & 16.34 \\ 
N821  & 39  & 3.3  & 38  & 3.1  & 36.0  & 7.02  & 5.76	 \\
N1052  & 41  & 2.5  & 35  & 2.2  & 35.4  & 7.42  & 5.93	 \\
N1380  & 36  & 3.9  & 28  & 3.4  & 28.0  & 10.99 & 8.90	 \\
N1404  & 33  & 3.6  & 18  & 2.5  & 17.9  & 15.14 & 12.87 \\
N3115  & 99  & 11.1 & 89  & 10.2 & 82.8  & 8.51  & 7.06	 \\
N3379  & 95  & 23.4 & 87  & 22.9 & 65.8  & 6.89  & 5.72	 \\
N3585  & 59  & 6.1  & 56  & 5.8  & 53.9  & 15.73 & 12.11 \\
N3923  & 105 & 10.8 & 83  & 9.8  & 89.6  & 26.36 & 21.61 \\
N4125  & 42  & 8.2  & 27  & 6.6  & 24.2  & 20.98 & 16.78 \\
N4278  & 177 & 16.3 & 160 & 14.9 & 154.2 & 6.71  & 5.24	 \\
N4365  & 244 & 23.9 & 213 & 22.5 & 201.1 & 18.87 & 16.04 \\
N4374  & 133 & 18.1 & 88  & 13.9 & 91.5  & 22.33 & 18.53 \\
N4382  & 52  & 13.5 & 44  & 12.3 & 33.1  & 25.14 & 19.11 \\
N4472  & 238 & 26.3 & 171 & 24.0 & 171.6 & 39.71 & 33.76 \\
N4552  & 94  & 10.4 & 68  & 7.6  & 70.7  & 9.07  & 7.52	 \\
N4636  & 123 & 12.2 & 82  & 9.1  & 83.6  & 11.39 & 9.22	 \\
N4649  & 236 & 15.0 & 149 & 10.7 & 168.5 & 28.97 & 24.62 \\
N4697  & 120 & 22.4 & 107 & 21.4 & 85.9  & 7.96  & 6.13	 \\
N5866  & 29  & 3.1  & 23  & 2.7  & 21.5  & 8.45  & 6.09	 \\
\hline
Total  & 2074& 240.0& 1626& 210.4& 1576.1& 316.6 & 259.4 \\ 
\hline
\end{tabular}
\end{center}
 (1) -- Galaxy name. (2) and (4) -- Number of all resolved X-ray point
 sources and sources brighter than $L_{\rm lim}$. (3) and (5) --
 Predicted number of CXB sources among (2) and (4). (6) -- Total
 number of LMXBs above $L_{\rm lim}$ after incompleteness correction
 and CXB subtraction. (7) and (8) -- Total $K_{\rm S}$-band
 luminosity and stellar mass (in units of $10^{10}$ $L_{\rm K,\odot}$
 and $10^{10}$ $M_{\odot}$) in the study field. 
\end{table}

\subsection{X-ray incompleteness correction}
\label{sec:icf}

The detection sensitivity of point sources varies throughout the
\chandra\ images. Reasons include the inhomogeneous level of
the diffuse X-ray emission in the galaxy, the deterioration of the PSF
at large off-axis angles, and the nonuniform exposure of an image in
which observations with different pointings are combined. To calculate
the point source detection sensitivity, we used the method and code
from \citet{Voss2006}, in which the detection method was inverted
using the local PSF, background, and exposure. The incompleteness
function $K(L)$ is computed as the fraction of pixels weighted by the
assumed spatial distribution of sources, in which detection
sensitivity is better (lower value) than the given luminosity. We
calculated $K(L)$ separately for the CXB sources and LMXBs, since the
CXB sources  have a flat distribution, while the field
LMXBs are expected to follow the $K_{\rm S}$-band light (for which we 
used the 2MASS Large Galaxy Atlas data \citep{Jarrett2003} -- see
Sect.~\ref{sec:kband}).   

To estimate the incompleteness-corrected number of LMXBs in the study
field in each galaxy, we did incompleteness correction for the number
of all resolved point sources, assuming $K(L)$ for the field LMXBs, and
then subtracted the corresponding number of CXB sources. This procedure 
is described by the equation 
\begin{eqnarray}
N_{\rm LMXB} = \sum_{L_{\rm i}>L_{\rm min}}^{L_{\rm max}} \frac{1}{K_{\rm LMXB}(L_{\rm i})} 
 - \nonumber \\
\int_{L_{\rm min}}^{L_{\rm max}} 4\pi D^2 \ \frac{dN_{\rm CXB}}{dL}\ \frac{K_{\rm CXB}(L)}{K_{\rm LMXB}(L)} \ dL,  
\label{eq:nlmxb}
\end{eqnarray}
where $4\pi D^2 dN_{\rm CXB}/{dL}$ equals $dN_{\rm CXB}/dS$, which
is the $\log(N)-\log(S)$ distribution of the CXB sources. We listed
the total number of LMXBs
above $L_{\rm lim}$, CXB subtracted and incompleteness corrected, in Table~\ref{tab:stat}.

\subsection{Near-infrared data analysis}
\label{sec:kband}

We calculated the stellar mass in the study field from near-infrared
data, using the $K_{\rm S}$ (2.16 $\mu$m) images from the 2MASS Large
Galaxy Atlas \citep{Jarrett2003} provided by the NASA/IPAC Infrared
Science archive. Most images are background subtracted, except for NGC
821, for which we obtained the background from adjacent regions. We also
removed the contamination of bright fore/background point sources from
images visually. The integrated, point source- and
background-subtracted count rate ($S$) was converted into calibrated
magnitudes with $m_{\rm K}$(mag) = KMAGZP-2.5log($S$), where KMAGZP is
the zero point magnitude for the $K_{\rm S}$-band given in the image
header. Thus the integrated $K_{\rm S}$-band luminosity was
calculated and then converted to stellar mass with the $K_{\rm
  S}$-band mass-to-light ratios (Table~\ref{tab:sample}) derived from 
  \citet{Bell2001} and with $B-V$ colors from RC3 catalog \citep{Dev1991}. We list the results in the
  Table~\ref{tab:stat}.

\begin{figure}
\resizebox{\hsize}{!}{\includegraphics[angle=0]{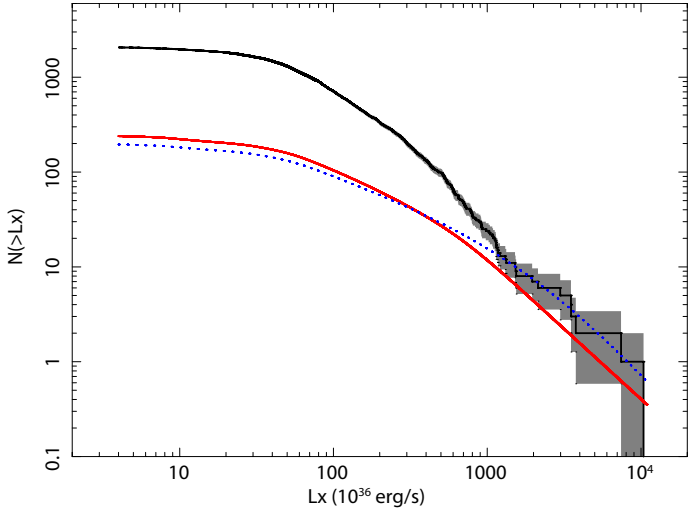}}
\caption{Observed cumulative distribution of all resolved point
  sources in all galaxies. The distribution is not corrected for
  incompleteness or the contribution of CXB sources. The shaded area
  shows $1\sigma$ Poissonian uncertainty. The thick solid and dotted
  lines show  the predicted distribution of CXB sources based on the
  $\log(N)-\log(S)$ from \citet{Geo2008} and \citet{Moretti2003},
  respectively.} 
\label{fig:cxb}
\end{figure}

\section{Average XLF and scaling relations for LMXBs}
\label{sec:xlf}

\subsection{XLF of compact sources and the CXB contribution}

Figure \ref{fig:cxb} presents the combined luminosity distribution of
all X-ray compact sources detected within the study fields of galaxies
and the predicted distribution of the CXB sources. It shows that
the CXB contribution is relatively unimportant below  $\log(L_{\rm X})=
39$, where it accounts for approximately 10\% of the
observed compact sources.  

Due to quick declining of the  LMXB XLF in the $\log(L_{\rm X})\sim
38.5-39$ range, an accurate account of the CXB contribution becomes crucial
at $\log(L_{\rm X})\gtrsim 39$. There is an apparent tail of the observed
source counts in this luminosity range, whose slope is similar to the
slope of the predicted distribution of CXB sources. Its normalization, 
however, is somewhat higher than predicted by the CXB $\log(N)-\log(S)$ 
from \citet{Geo2008}. Quantitatively, we detected 24 sources above $10^{39}$ erg/s, 
while 11.8 background AGN in these fields is predicted, based on \citet{Geo2008}. 
The Poissonian distribution predicts a low probability of $\sim 1.2\cdot 10^{-3}$ for such a
deviation solely due to random fluctuations. We also checked the
predictions of the CXB $\log(N)-\log(S)$ determined by \citet{Moretti2003}
and found that they can fully account for the observed bright sources,
as shown by the dotted line in Fig.~\ref{fig:cxb}. In computing this
prediction, we used the soft band (0.5-2 keV) counts and converted
them to the $0.5-8$ keV band, as described in \citet{Zhang2011}. 

Comparing the two predictions, the following remarks should be made. The more recent 
work of \citet{Geo2008} is based on a larger sample of sources detected in a larger 
number of \chandra\ surveys, and its results are in good agreement with another recent 
study by \citet{Kim2007}. Moreover, converting the soft band flux from \citet{Moretti2003} 
into 0.5-8 keV highly depends on the assumed spectrum of CXB sources, for example, 
changing the photon index from 1.4 to 1.9 makes $\sim1.5$ times difference. In addition, 
the numbers of bright sources detected outside the D25 region of galaxies tends to
be in agreement with the prediction of the $\log(N)-\log(S)$ by \citet{Geo2008} (see below). 
For these reasons, we decided to accept \citet{Geo2008} as our default CXB model.

\begin{figure}
\resizebox{\hsize}{!}{\includegraphics[angle=270]{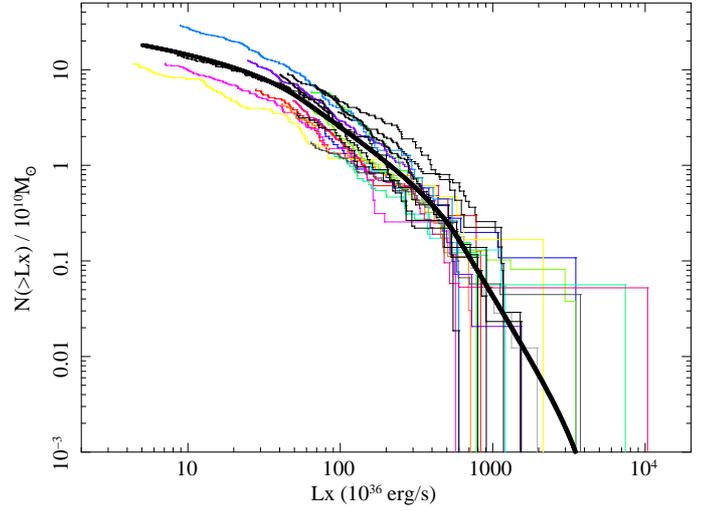}}
\caption{Cumulative XLFs of LMXBs in galaxies of
  our sample. They are CXB subtracted, 
  incompleteness corrected, and normalized to the stellar mass of the 
  host galaxy. They are plotted above corresponding $L_{\rm lim}$ of 
  each galaxy. The solid line is the average XLF of LMXBs in nearby 
  galaxies from \citet{Gilfanov2004}.} 
\label{fig:cumuxlf}
\end{figure}

\subsection{Average XLF of LMXBs}
\label{sec:cxlf}

The CXB-subtracted and incompleteness-corrected cumulative XLFs of compact X-ray sources in each galaxy are
plotted in Fig.~\ref{fig:cumuxlf}. As LMXBs are almost 
the only type of compact X-ray source in early-type galaxies capable 
of emitting at the $\log(L_{\rm X})\gtrsim 36$ luminosity level, the 
distributions shown in Fig.~\ref{fig:cumuxlf} can be regarded as luminosity
functions of LMXBs in these galaxies. The XLFs have
been normalized to unit stellar mass in the study field. It is clear
that all the XLFs have a similar shape, which is broadly consistent
with the average XLF of LMXBs in nearby galaxies
obtained by \citet{Gilfanov2004} (plotted with the thick line in the
figure). On the other hand,  a notable scatter of more than a factor
of two exists for the normalization, which is a manifestation of the
scatter in the LMXB-stellar mass relation, as discussed below.

\begin{figure}
\resizebox{\hsize}{!}{\includegraphics[angle=0]{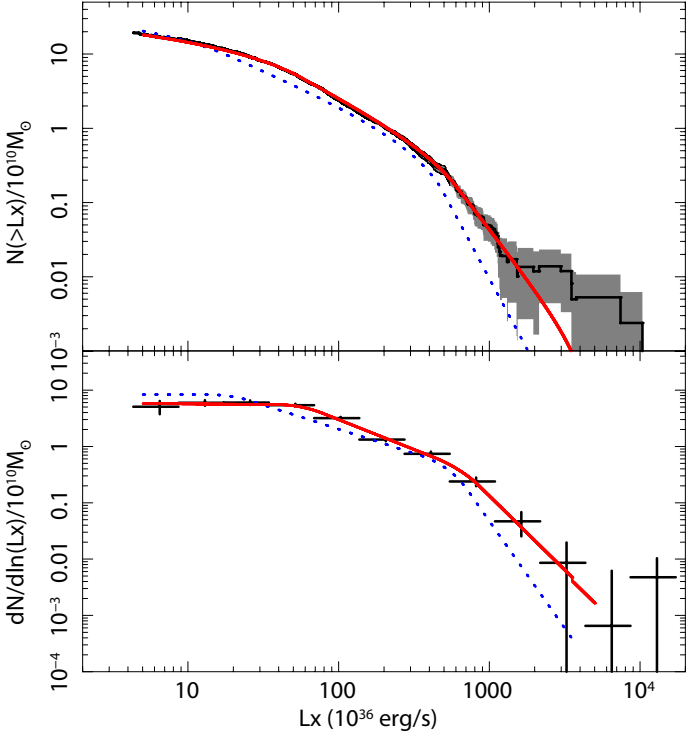}}
\caption{Combined luminosity distribution of LMXBs in our
  sample in cumulative (upper panel) and differential (lower panel) forms. 
  The distributions are CXB subtracted and incompleteness
  corrected, as described in the text. The shaded area in the upper
  panel indicates the $1\sigma$ Poissonian uncertainty. The solid
  lines show the best-fit model with two breaks, the dotted lines show
  the average LMXB XLF from \citet{Gilfanov2004}. See Sect.~\ref{sec:ulx} 
  for discussion and caveats of the high-luminosity tail.} 
\label{fig:comxlf}
\end{figure}

To construct combined XLF of all galaxies with different detection
sensitivity, we followed the method described in
\citet{Zhang2011}. The cumulative and differential forms of our XLF
are plotted in Fig.~\ref{fig:comxlf}. We fitted the combined XLF with
the template introduced in \citet{Gilfanov2004}: 
\begin{eqnarray}
\frac{dN}{dL_{36}}=\left\{ \begin{array}{ll}
\renewcommand{\arraystretch}{3}
K_1 \left(L_{36}/L_{\rm b,1}\right) ^{-\alpha_1},	
			& \mbox{\hspace{1.9cm} $L_{36}<L_{\rm b,1}$}\\
K_2 \left(L_{36}/L_{\rm b,2}\right)^{-\alpha_2},	
			& \mbox{\hspace{1.0cm} $L_{\rm b,1}<L_{36}<L_{\rm b,2}$}\\
K_3 \left(L_{36}/L_{\rm cut}\right)^{-\alpha_3},	
			& \mbox{\hspace{1.0cm} $L_{\rm b,2}<L_{36}<L_{\rm cut}$}\\
0			& \mbox{\hspace{2.0cm} $L_{36}>L_{\rm cut}$}\\
\end{array}
, \right.
\label{eq:uxlf}
\end{eqnarray}
where $L_{36}=L_{\rm X}/10^{36}$ erg/s and normalizations $K_{1,2,3}$ are related by:
\begin{eqnarray}
K_2=K_1 \left(L_{\rm b,1}/L_{\rm b,2}\right)^{\alpha_2}\nonumber, \\
K_3=K_2 \left(L_{\rm b,2}/L_{\rm cut}\right)^{\alpha_3}\nonumber. 
\end{eqnarray}
The value of the high-luminosity cut-off was fixed at $L_{\rm
  cut}=5\cdot10^4$. We performed maximum-likelihood fitting to the
unbinned data. Our best-fit parameters  with 1$\sigma$ errors are: 
$\alpha_1=1.02^{+0.07}_{-0.08}$, $\alpha_2=2.06^{+0.06}_{-0.05}$, $\alpha_3=3.63^{+0.67}_{-0.49}$,
$L_{\rm b,1}=54.6^{+4.3}_{-3.7}$ and $L_{\rm b,2}=599^{+95}_{-67}$. 
The normalization is $K_1=1.01\pm0.28$ per $10^{11}$ M$_{\odot}$.

The combined XLF obtained in this study is broadly consistent with
the average LMXB XLF obtained by \citet{Gilfanov2004} (cf.\ dotted
line in Fig.~\ref{fig:comxlf}). The XLF of the sources in our sample
appears to be somewhat flatter in the bright end of $\log(L_{\rm X})\gtrsim 38.5$,
having more luminous sources. As evident in Sect.~\ref{sec:xlf_age}, 
this is related to the large fraction of younger galaxies in our sample. Additionally, 
there is a rather peculiar tail of luminous sources above $\log(L_X)\ga 39$, 
which will be discussed in Sect.~\ref{sec:ulx}.

\subsection{Scaling relations for LMXBs}
\label{sec:lmxbms}

\begin{figure}
\resizebox{\hsize}{!}{\includegraphics[angle=270]{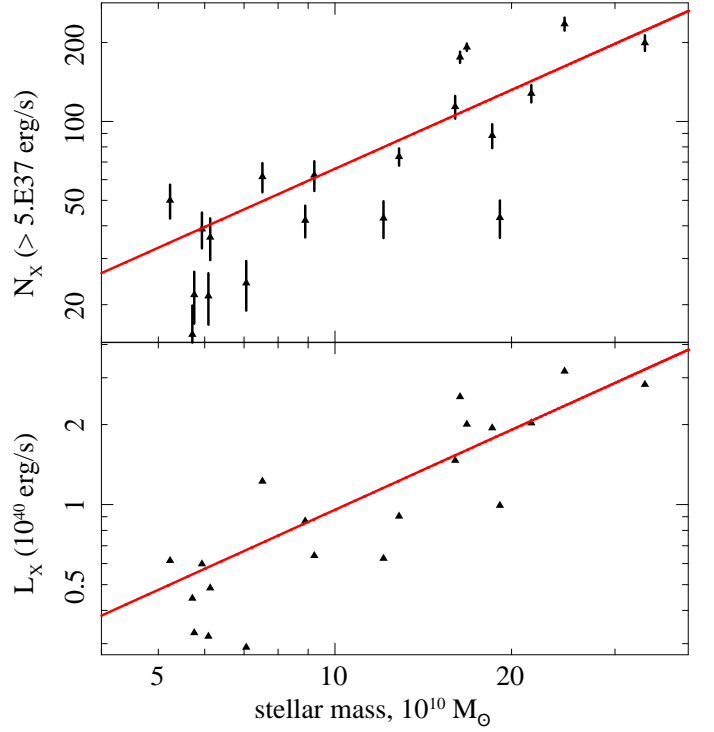}}
\caption{Relation of the total number (upper panel) and luminosity
  (lower panel) of LMXBs in the study field in each galaxy 
  with the stellar mass. The solid lines show the best fittings.} 
\label{fig:scale}
\end{figure}

By definition, the present sample is not designed for detailed analysis 
of scaling relations of LMXBs with stellar mass, since selected galaxies occupy 
a rather narrow range of masses (less than a factor of $\la 7$). The XLF analysis 
presented above suggests that our data is generally consistent with the $N_X-M_*$ 
and $L_X-M_*$ dependence obtained previously. This is analyzed below.

The total numbers and collective luminosities of LMXBs that are more luminous than 
$5\cdot10^{37}$ erg/s (after subtracting the CXB contribution and correcting for 
the incompleteness, Eq.(\ref{eq:nlmxb})) for all the galaxies are plotted against 
stellar mass in Fig.~\ref{fig:scale}. The combined XLF of our sample gives the
values of $N_{\rm X}/M_*$ = 54.3 and $L_{\rm X}/M_*$ =
$8.5\cdot10^{39}$ erg/s per $10^{11}$ M$_{\odot}$. The fit to the data of individual 
galaxies shown in the scatter plots in Fig.~\ref{fig:scale} gives 
$N_{\rm X}/M_*$ = 66.0 and $L_{\rm X}/M_*$ = $9.6\cdot10^{39}$ erg/s per $10^{11}$ M$_{\odot}$. 
These values are $\sim$1.4 -- 1.7 times larger than those obtained by \citet{Gilfanov2004} (37.8 
and $5.8\cdot10^{39}$ erg/s). This is caused by different age distributions and GC 
contents of the two samples. The best-fit relations are shown by the solid lines in the
plot. The $rms$ deviation of the points from best-fit relations are 0.19 dex for 
$N_{\rm X}$ and 0.16 dex for $L_{\rm X}$. The observed scatter is likely caused 
by the large spread in the galaxy ages and (related to it) variation in the 
contribution of binaries formed dynamically in GCs. These effects are 
further discussed in the next section.

\section{Dependence on stellar age and GC content}
\label{sec:dependence}

\subsection{Correlation of LMXBs with age and GC content of the host galaxy}
\label{sec:correlation}

To characterize the number of LMXBs per unit stellar mass in each galaxy,
we used the following quantity
\begin{equation}
f_{\rm XLF}=\frac{N_{\rm X}(L>L_{\rm lim})-N_{\rm CXB}(L>L_{\rm lim})}{M_{*}\times\int_{L_{\rm lim}}F(L) K_{\rm LMXB}(L) dL},
\label{eq:fXLF}
\end{equation} 
where $N_{\rm X}(L>L_{lim})$ and $N_{\rm CXB}(L>L_{lim})$ are
the numbers of detected X-ray sources and predicted CXB sources
(Table~\ref{tab:stat}), $F(L)$ is the best-fit differential XLF normalized to $10^{11}~M_\odot$, $M_*$ is the stellar mass in units of $10^{11}~M_\odot$,
and $K_{\rm LMXB}(L)$ is the incompleteness function of LMXBs in the given
galaxy. The so-defined  \fxlf\ is the specific (per $10^{11}~M_\odot$) XLF normalization
computed from the number of resolved LMXBs above $L_{\rm lim}$, assuming the average 
XLF shape of the resolved sources, modified by the point-source detection incompleteness. 
For the average XLF, we used the  best-fit  model from Sect.~\ref{sec:xlf}. 
With this definition, the \fxlf\ equals 1 corresponds to 5.4 LMXBs  with luminosity $L_X>5\cdot10^{37}$ 
erg/s per $10^{10}~M_\odot$.

Obviously, \fxlf\ has the advantage that all detected X-ray sources above 0.6 incompleteness level are involved 
in the calculation. The disadvantage is that it relies on the assumption that the XLF shape does not change from 
galaxy to galaxy. However, it is known that the XLFs of GC and field LMXBs are different \citep{Zhang2011}, 
as well as the LMXBs in young and old galaxies \citep{Kim2010}. The systematic bias can be further amplified by the 
fact that different galaxies have different values of $L_{lim}$. To investigate the importance of these effects, 
we computed \fxlf\ for several galaxies using the average XLFs for young and 
old galaxies (see the next section) and found that they differ by no more than $\sim 30\%$. This is 
insignificant given the scatter of the points and the amplitude of the correlations found below. 
As a further check, we recomputed \fxlf\ in the luminosity range of  $5\cdot10^{37}$ to $5\cdot10^{38}$ 
erg/s where both effects are minimal. We did not find any systematic changes in our results.  

\begin{figure*}
\hbox{
\resizebox{!}{0.365\hsize}{\includegraphics[angle=90]{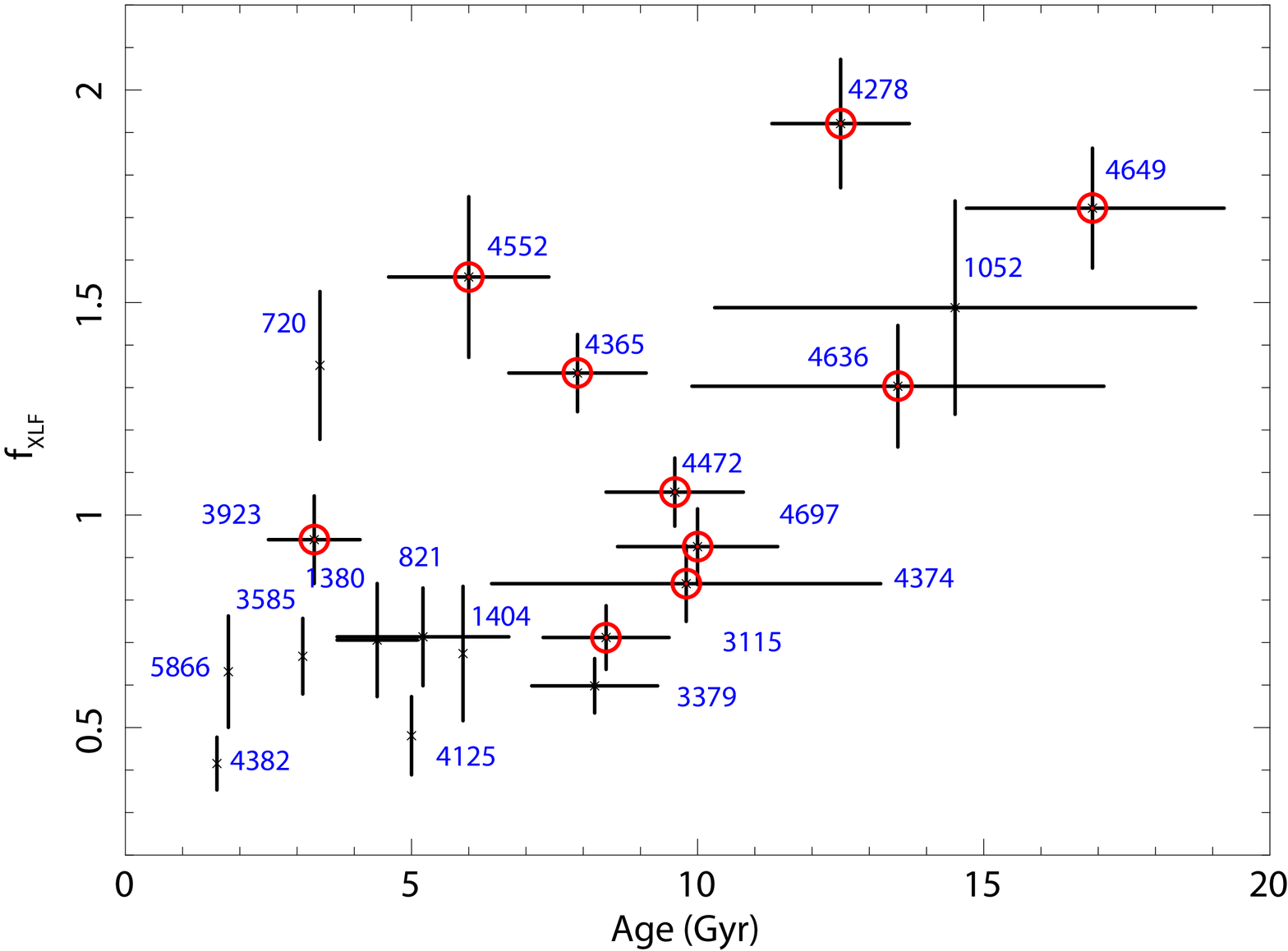}}
\resizebox{!}{0.37\hsize}{\includegraphics[angle=90]{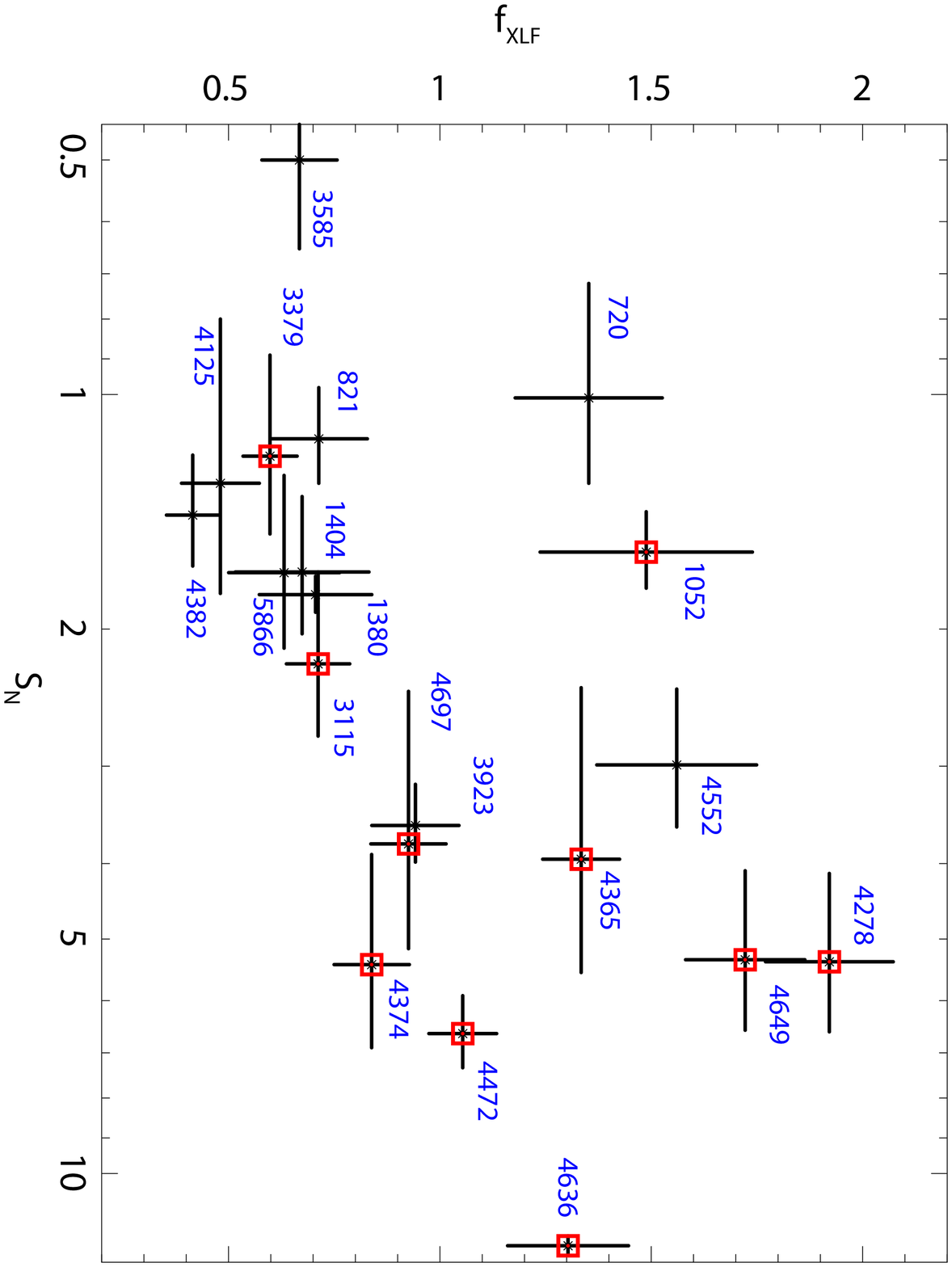}}
}
\caption{Correlation of  the XLF normalization, \fxlf\ (Eq.~\ref{eq:fXLF}), with stellar age (left) and \sn\ (right). Circles mark galaxies with $S_{\rm N}>2.0$ (left panel), and squares mark galaxies with the stellar age $>$6 Gyr (right panel).} 
\label{fig:n_age_sn}
\end{figure*}

We plot \fxlf\ versus stellar age and \sn\ in Fig.~\ref{fig:n_age_sn}. 
Both plots show moderate trends that the \fxlf\ increases with the stellar age of the host galaxy 
and its GC content. Despite rather large scatter of the points, the Spearman rank-order 
correlation test gives the null hypothesis probability of $p=0.003$ for the age- and much 
larger value of $p=0.017$ for the \sn-dependence. These numbers indicate a moderately significant 
correlation of $\sim 2.5-3\sigma$. 

The correlation of \fxlf\ with \sn\ is similar to the one found in previous studies \citep[e.g.,][]{Humphrey2008,Boroson2011}. 
The presence of such correlation was interpreted as evidence suggesting that a significant part 
(if not all) of the LMXB population, including the field sources, was formed in GCs 
and subsequently expelled into the field. The existence of the equally strong correlation of the 
specific LMXB frequency with the age of the stellar population suggests that this interpretation 
is not unique and the more important correlation may be with the stellar age. 

In order to investigate this further, we fit the data with a two-parameter linear model in the form 
$f_{\rm XLF}=a\times t+b\times S_{\rm N} + c$. Using $\chi^2$ minimization, we found the following values
of best-fit parameters: $a=0.044\pm0.008$, $b=0.049\pm0.012$, $c=0.385\pm0.047$, with a very 
large value of $\chi^2=129.7$ for 17 d.o.f. With these best-fit values the contribution of the 
age term is about two times larger than the contribution of the \sn\ term, suggesting that the more 
important parameter is the age, rather than the GC content of the host galaxy. 
However, because of the rather large dispersion of the points and correspondingly large value 
of $\chi^2$, any firm conclusion is premature.  

The average values of \fxlf\ for galaxies younger and older than 6 Gyrs are $0.75\pm0.08$ and  
$1.08\pm0.06$, respectively. The statistical significance of the difference between these two values 
is $\approx 3.3\sigma$, in agreement with the Spearman test results. The total number of LMXBs per
unit stellar mass above $5\cdot10^{37}$ erg/s in the young galaxy sample is $4.17\pm0.27$, which is $\sim$2/3 
of that in the old ($6.27\pm0.26$). The prediction from the average XLF is 5.4.

\subsection{XLFs of young and old galaxies}
\label{sec:xlf_age}

In order to investigate the age dependence of the LMXB XLFs, we
divided galaxies into young and old groups using their median age (6 Gyrs) as a boundary. 
Each group contains ten galaxies, which are marked correspondingly in Table~\ref{tab:age}. 
The study regions in young and old galaxies cover a total solid angle of 125.9 and 251.6 arcmin$^{2}$ 
respectively, with a total stellar mass of 1.28 and $1.32\cdot10^{12}M_\odot$.  

\begin{figure}
\resizebox{\hsize}{!}{\includegraphics[angle=0]{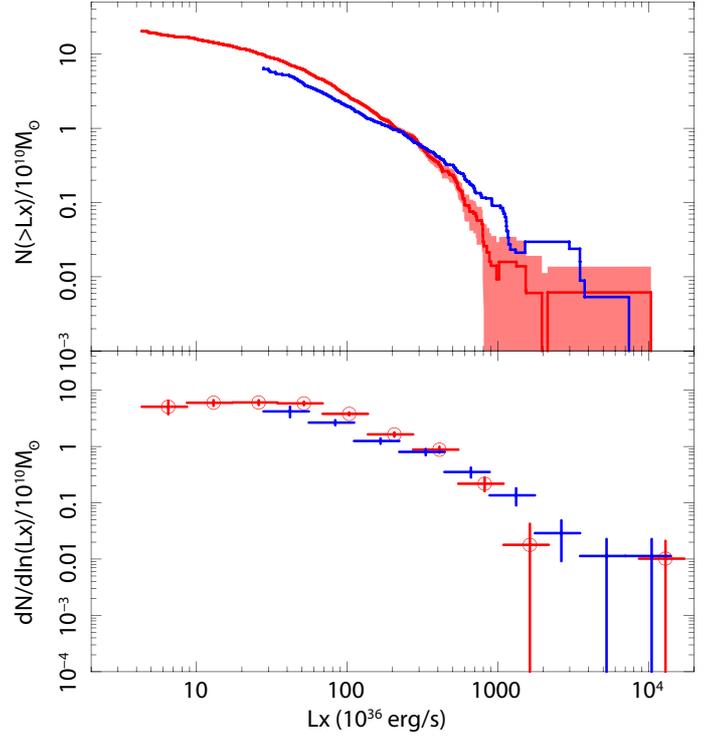}}
\caption{XLFs of LMXBs in young and old galaxies
in cumulative (upper panel) and differential (lower panel) forms. The data for old galaxies 
(red in the color version of the plot) is marked by circles in the lower panel and is 
surrounded by the shaded area showing  the $1\sigma$ Poissonian uncertainty in the upper panel. 
Statistical uncertainty for young galaxies has comparable amplitude.}
\label{fig:xlfage}
\end{figure}

We constructed combined XLF of each group and plotted them in
Fig.~\ref{fig:xlfage}. In general, older galaxies have deeper
\chandra\ observations that have reached a sensitivity of
$\sim 5\cdot10^{36}$ erg/s, while the young group has a sensitivity
of $\sim 3\cdot10^{37}$ erg/s. The overall shape of the XLF for
young galaxies is flatter than that of the old ones. This behavior
is in agreement with findings of \citet{Kim2010}.  

\begin{figure}
\resizebox{\hsize}{!}{\includegraphics[angle=0]{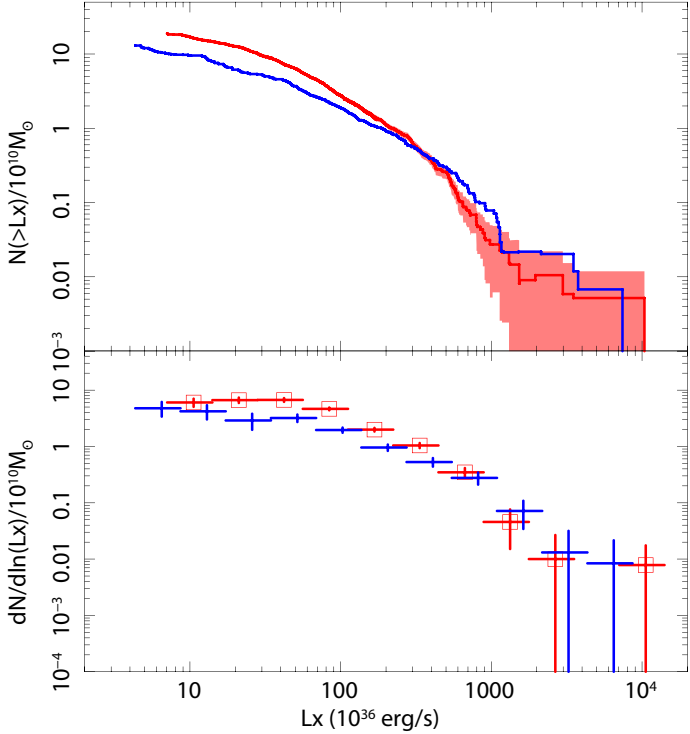}}
\caption{XLFs of LMXBs in galaxies with high
and low \sn\ (see text for details) in cumulative (upper panel) and differential (lower panel) forms. 
The data for high-\sn\  galaxies (red in the color version of the plot) is marked by 
squares in the lower panel and is surrounded by the shaded area showing  the $1\sigma$ 
Poissonian uncertainty  in the upper panel. 
Statistical uncertainty for low-\sn\ galaxies has comparable amplitude.} 
\label{fig:xlfsn}
\end{figure}

Similarly, we use the median value of the \sn\ distribution ($S_{\rm N}=2.0$) to 
divide galaxies into GC-rich and GC-poor subgroups. The resulting XLFs are shown in 
Fig.~\ref{fig:xlfsn}. As expected from the age dependence of the XLF and the general
correlation between age and \sn, the combined XLF of GC-rich galaxies is steeper than 
that of the GC-poor ones. However, unlike XLFs of young and old galaxies, they appear 
to be similar in the bright end $\log(L_{\rm X})\gtrsim 38.5-39$.

\begin{table*}
\begin{center}
\caption{Sources brighter than $10^{39}$ erg/s.} 
\label{tab:ulx}
\begin{tabular}{rllrll}
\hline
\hline
\vspace{0.1cm}
Galaxy & RA(J2000)     & DEC(J2000)     & Luminosity      & D25 & opt. \\   
\hline   
       &               & Young Galaxies ($<6$ Gyrs) &                 &     &      \\   
\hline 
N720   & +01:52:56.50  &  -13:43:47.77  &  $1.04\pm0.08$  & 0.55 & GC  \\  
       & +01:53:06.43  &  -13:45:40.58  &  $1.08\pm0.08$  & 0.86 & --  \\
       & +01:52:55.82  &  -13:43:50.90  &  $1.14\pm0.08$  & 0.67 & --  \\
       & +01:52:59.39  &  -13:43:57.28  &  $1.17\pm0.08$  & 0.19 & GC  \\
       & +01:53:01.12  &  -13:44:19.53  &  $1.52\pm0.08$  & 0.10 & --  \\	
N1380  & +03:36:26.56  &  -34:56:58.96  &  $1.09\pm0.09$  & 0.73 & --  \\
       & +03:36:25.25  &  -34:59:18.09  &  $3.51\pm0.04$  & 0.47 & --  \\  
N1404  & +03:38:51.99  &  -35:35:59.93  &  $1.14\pm0.07$  & 0.20 & --  \\  
       & +03:38:54.78  &  -35:35:00.96  &  $1.21\pm0.07$  & 0.57 & --  \\  
N3923  & +11:50:58.65  &  -28:49:13.16  &  $1.31\pm0.08$  & 0.38 & --  \\
       & +11:51:09.54  &  -28:48:00.67  &  $2.98\pm0.12$  & 0.68 & --  \\
       & +11:51:06.22  &  -28:46:49.91  &  $3.50\pm0.13$  & 0.65 & Q   \\ 
N4125  & +12:08:07.46  &  +65:10:28.61  &  $7.41\pm0.23$  & 0.05 & --  \\  
N4382  & +12:25:20.32  &  +18:13:01.41  &  $1.12\pm0.09$  & 0.58 & --  \\  
       & +12:25:17.17  &  +18:13:46.52  &  $3.76\pm0.18$  & 0.93 & Q   \\  
N4552  & +12:35:45.77  &  +12:33:02.46  &  $1.14\pm0.06$  & 0.61 & GC  \\  
       & +12:35:41.22  &  +12:34:51.43  &  $1.18\pm0.06$  & 0.62 & GC  \\  
\hline
       &               & Old Galaxies  ($>6$ Gyrs) &      &  &         \\
\hline
N3379  & +10:47:50.01  &  +12:34:56.77  &  $2.14\pm0.03$  & 0.04 & --  \\  
N4365  & +12:24:26.36  &  +07:16:53.55  &  $1.53\pm0.05$  & 0.71 & --  \\  
N4374  & +12:25:11.92  &  +12:51:53.81  &  $10.38\pm0.16$ & 0.74 & Q   \\  
N4472  & +12:29:41.00  &  +07:57:44.46  &  $1.96\pm0.08$  & 0.62 & GC  \\  
       & +12:29:34.46  &  +07:58:51.63  &  $1.33\pm0.06$  & 0.79 & G   \\  
       & +12:29:42.33  &  +08:00:07.96  &  $1.02\pm0.05$  & 0.25 & GC  \\  
N4649  & +12:43:46.90  &  +11:32:34.19  &  $1.54\pm0.06$  & 0.48 & --  \\  

\hline
\end{tabular}
\end{center}
Columns are the host galaxy name, coordinates, luminosity in units of
$10^{39}$ erg/s, offset from the center in units of  the D25 radius,
optical counterpart from NED (Q: Quasi-stellar object, GC: globular cluster, 
G: galaxy).   
\end{table*}

\section{Nature of (ultra-) luminous X-ray sources}
\label{sec:ulx}

After subtracting the contribution of background sources, a rather peculiar tail of luminous sources 
with $\log(L_X)\ga 39$ remains (Fig. \ref{fig:comxlf}). Although the CXB contribution is insignificant 
throughout most of the luminosity range, its correct subtraction is critical for 
establishing the nature of bright sources. In total,  we detected 24 compact sources with luminosity 
exceeding $10^{39}$ erg/s. Subtracting from this number the 11.8 background AGN predicted from the
\citet{Geo2008} CXB $\log(N)-\log(S)$, we obtain that $\sim 12-13$ sources should be associated with 
galaxies from our sample. Similarly, among sources with $L_{\rm X}>2\cdot 10^{39}$ erg/s (7 sources 
observed, 4.3 background AGN predicted), $\sim 2-3$ sources are expected to be associated with galaxies. 
Assuming Poissonian distribution, these numbers correspond to moderately low probabilities of being a result
of pure statistical fluctuations: $\sim 1.2\cdot 10^{-3}$ and 0.14 for the two luminosity ranges, 
suggesting that the luminous sources may indeed be X-ray binaries. 

On the other hand, it is well known that the CXB source counts produce somewhat different results in 
different sky fields due to cosmic variance. Moreover, the slope of the bright tail of the luminosity 
distribution of all compact sources in Fig.~\ref{fig:cxb} is similar to the slope of the CXB $\log(N)-\log(S)$. 
It is therefore possible that the tail of bright sources  in Fig. \ref{fig:comxlf} is due to background AGN, which is unaccounted for due to cosmic variance. The amplitude of CXB source density variations depends primarily on the
considered angular scales and decreases as the solid angle increases. Because we combined data from 20 galaxies distributed over the
extragalactic sky, we do not expect cosmic variance to be particularly strong in the combined LMXB XLFs shown 
in Figs.~\ref{fig:comxlf} and~\ref{fig:xlfage}.  In order to investigate this further, we checked the source 
numbers outside the D25 but within $\sim 10\arcmin\times10\arcmin$ around the aim point of the \chandra\ 
observation. In total 41 sources with $\log(L_{\rm X})>39$ were detected, while the CXB $\log(N)-\log(S)$ 
of \citet{Geo2008} predicts 49.8 background AGN. These two numbers are consistent within $\sim 1.4 \sigma$,  
which suggests that the effect of cosmic variance is not very strong. Moreover, the observed local CXB source 
density is $\sim 20\%$ lower than predicted, i.e., in computing the LMXB XLF shown in Fig. \ref{fig:comxlf}, 
the CXB contribution may be somewhat over-subtracted.

Furthermore, we checked the luminous, $L_X>10^{39}$ erg/s sources individually. In particular, we looked up 
their identifications in the NASA Extragalactic Database and checked their location relative to the D25 radius 
of the host galaxy. The results are presented in Table~\ref{tab:ulx}. In the table, we list young and old galaxies 
separately, using as before the median age of 6 Gyrs as the boundary. Among the 24 luminous sources, 17 are 
detected in the young galaxies and 7 in the old ones, with the CXB source predictions being 5.8 and 6.0, respectively. 
Taken at face value, these numbers suggest that all luminous sources are located in young galaxies. On the 
other hand, the luminous sources are evenly distributed between GC-rich (12 sources detected, 7.2 background AGN 
expected) and GC-poor galaxies (12 sources detected, 4.6 background AGN expected).

According to the NED, four of the luminous sources are confirmed background AGN, including  
the brightest sources with $L_X\sim 10^{40}$ erg/s located in NGC 4374. As expected, this 
number is significantly smaller than the predicted number of $\approx 12$ AGN, due to incompleteness 
of the NED. Interestingly, six luminous sources are located in GCs, two of which are 
in the old galaxy sample. All six sources have luminosities $\sim 1-2\cdot10^{39}$ erg/s, which makes them 
plausible black hole candidates. Both GC sources in NGC 720 were found by \citet{Jeltema2003}. Both sources in NGC 4472, (+12:29:42.33,+08:00:07.96) and (+12:29:41.00,+07:57:44.46) were previously reported  by \citet{Mac2003, Mac2011}, respectively. 
Of the two sources in NGC 4552, one (+12:35:45.77,+12:33:02.46) was previously known 
\citep{Xu2005}, while the other (+12:35:41.22,+12:34:51.43) is,  to our knowledge, 
a new identification.

The above analysis suggests that the XLF of compact sources in young galaxy sample extends significantly 
beyond $10^{39}$ erg/s, with the specific frequency of luminous sources $\approx 8.8\pm3.2$ sources per 
$10^{12}~M_\odot$. Although the numbers of detected sources and uncertainties of the CXB subtraction 
do not allow us to constrain its shape, it appears to be a smooth extension of the LMXB XLF observed  
at lower luminosities. The luminosities of these sources, up to $\sim 3-7\cdot 10^{39}$ erg/s, are 
compatible with, or exceed only slightly, the Eddington limit for a $\sim 10~M_\odot$ black hole. We therefore 
believe that most likely these sources are stellar mass black holes accreting from a low or intermediate 
mass companion. In the case of an evolved companion, such systems may be transient sources 
\citep{King1997,Piro2002}. In order to check this possibility, we investigated the variability of those luminous 
sources for which multiple observations were available. Although bright sources show significant variability, we did not find any evidence for transient behavior up to a factor of $\sim 5$.

\section{Caveats and uncertainties}
\label{sec:caveats}

The accuracy of our conclusions obviously depends on the accuracy of the quantities used to characterize stellar ages and GC content of galaxies. However, both quantities are subject to several, currently unavoidable, uncertainties.  In the following subsections, we discuss these uncertainties in detail. 

\subsection{Stellar age determinations}
\label{sec:caveats_age}

Stellar age determinations are based on the comparison of the strengths of the  absorption lines of age-sensitive elements with predictions from the stellar synthesis models. In the core of their algorithms lies the assumption of SSP.
The SSP assumption is an obviously oversimplified picture of the stellar populations  in early-type galaxies. This is especially true when the entire galaxy is characterized by a single value of the stellar age. In this case, the problem can be addressed considering the spatially resolved stellar age maps, as discussed below. It is more complicated when a small fraction of the young population is mixed with the older underlying population, thus biasing the SSP-equivalent age towards younger ages \citep[e.g.,][]{Salim2010,Marino2011}. In this case, more sophisticated assumptions regarding the stellar population content have to be made. At present, such analysis is not available for a sufficiently large number of galaxies for statistically meaningful study. Also, the complexity of the answer obtained in this case would not allow an analysis in terms of simple age--\fxlf\ plots. 

More ``technical" issues also affect the outcome of age determination in each particular galaxy. 
These include, for example, differences in the correction 
for the ionized gas emission, in the selection of absorption lines used for fitting the spectral population models, and in the choice of  libraries of stellar population synthesis models \citep[see, e.g.,][]{Annibali2007}. As a result, there is a spread of values obtained by different authors that is sometimes rather large. We tried to minimize the effect of these issues by assigning priorities to different age determinations, as described in the Sect.~\ref{sec:age}.

Many of the age measurements analyze the spectrum of the very small central 
region of the galaxy, usually corresponding to $r_{\rm e}/8$, where $r_{\rm e}$ is the effective radius 
of the galaxy with a typical size of $30^{\prime\prime}$. As LMXBs are rare objects, their numbers inside 
$r_{\rm e}/8$ detected in a typical \chandra\ 
observation of a typical galaxy are by far too insufficient for any statistically meaningful analysis. 
On the contrary, to increase their numbers, LMXBs are collected from a region whose size 
is comparable with the D25 diameter. Moreover, the central region of the size of $\sim$ few arcsec 
is usually excluded from the X-ray point source analysis in order to exclude from the analysis the central (weak) AGN, 
peaked diffuse emission, and source confusion. Thus, the LMXBs and age measurements are 
inevitably performed in geometrically different, sometimes barely overlapping regions of the galaxy. 
Obviously, such analysis requires an assumption of the homogeneity of the stellar population, which 
may not be fulfilled in all cases. The statistical study of \citet{Tortora2010} investigated a sample 
of $\sim$50000 nearby galaxies with the Sloan Digital Sky Survey data. Their results show no age gradient
for massive early-type galaxies ($>10^{10}M_\odot$) with a central age older than 6 Gyr, while for the 
ones with a central age younger than 6 Gyr, age gradients can be as big as $\nabla_{\rm age}\sim0.4$ 
(2.5 times difference). Based on the spatially resolved age maps delivered by the SAURON project, \citet{Kuntschner2010} came to a similar conclusion. This result suggests that the ages of old galaxies in our sample should be on average sufficiently reliable, whereas an additional scatter of a factor of $\sim 2$ along the age axis may be possible for younger galaxies. Evidence for such behavior can possibly be seen in Fig.\ref{fig:n_age_sn}.

To investigate this further, we searched for individual studies of the age gradients galaxy by galaxy.  The six galaxies included in the SAURON sample (NGC 821, NGC 3379, NGC 4278, NGC 4374, NGC 4382, and NGC 4552) do not have significant age gradients. 
Based on the long-slit spectroscopy results, we found evidence for complicated age structure in the following galaxies. NGC 720 was suggested to have been formed by a merger of an old (5-13 Gyr) small-scale spheroid and a younger (2.5-5 Gyr) large-scale disk component \citep{Rembold2005}, meaning that the average stellar age of NGC 720 should be older than 3 Gyr value adopted in this paper. NGC 4125 was found to have experienced a recent merger event: 
thus young stellar components likely exist in this galaxy \citep{Pu2010}. NGC 4365 is  likely to be older than 7.9 Gyr, as the results of \citet{Davies2001} suggest that the decoupled core and the main body of the galaxy have the same age of $\sim$14 Gyr. With these numbers, the data points for NGC 720 and NGC 4365 should shift to the right in the left-hand panel of Fig.~\ref{fig:n_age_sn}, while NGC 4125 may shift to the left. Such a correction will reduce the scatter on this plot.
No such detailed studies could be found for other galaxies from our sample. This analysis confirms that at least a part of the scatter on the age -- \fxlf\ plot is due to complexities of the age structure of the early-type galaxies. 

The above discussion suggests a promising venue for  future studies, based on the  comparison of the spatial distribution of compact sources with  spatially resolved stellar age maps. This is similar to the analysis conducted by \citet{Sht2007} for high-mass X-ray binaries, based on the comparison of their spatial distribution with spatially resolved star-formation history in the Small Magellanic Cloud. The outcome of such an analysis would be the full-time dependence of the LMXB population passed from the star-formation event.

\subsection{Specific frequency of GCs (\sn) }
\label{sec:caveats_gc}

The main complication in determining  the GC content is the lack of  uniform optical data suitable for this purpose, in particular individual GC lists for all galaxies. For this reason, we had to rely on the published total numbers of GC candidates detected in each galaxy to calculate the \sn. Since LMXBs are collected from the D25 region of each galaxy, GCs in the same region should be counted. However, published numbers often referred to different regions. For example, the HST data typically covered regions smaller than D25, while the data from ground-based telescopes often had a larger field of view. Because we compute V-band luminosity for the same region where GCs are counted, this would not matter if the spatial distribution of GCs followed that of stellar light. However, GCs are known to have a broader distribution than stars \citep{Bassino2006}, so the value of \sn\ depends on the spatial region over which it was computed.  
For example, the \sn\ values measured from the HST data 
\citep[e.g.,][]{Humphrey2009,Peng2008} is on average  smaller than the ones determined by the ground-based telescopes (see references in Table \ref{tab:sample}). 

The picture is further complicated by the fact that red (metal-rich) 
GCs  are more concentrated toward the galaxy centers, more accurately tracing the stellar population, while the blue (metal-poor) GCs are often more extended \citep[e.g.,][]{Bassino2006}. Since there are approximately three times as many LMXBs in red GCs as in blue GCs \citep[e.g.,][]{Bellazzini1995}, an accurate selection of the spatial regions and, ideally, a separation of the red and blue GCs are important. 

Our ability to take these complications into account is limited due to the lack of uniform optical data for the GC systems in the galaxies in our sample. These complications introduce scatter in our data, which currently cannot be eliminated.

\subsection{Correlation between stellar age and \sn}
\label{sec:age_sn}

\begin{figure}
\resizebox{\hsize}{!}{\includegraphics[angle=90]{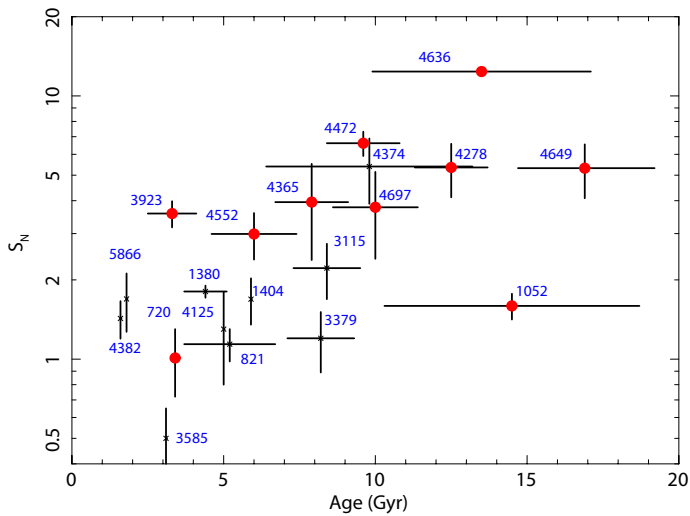}}
\caption{Correlation of \sn\ with stellar age. Filled circles mark galaxies with larger \fxlf.}
\label{fig:snage}
\end{figure}

In our sample, there is a strong correlation between stellar age and \sn, as illustrated by 
Fig.~\ref{fig:snage}. Although we did not find any report about such a correlation in the literature, its 
existence is not surprising because more GCs are expected to be found in older galaxies as a result of more massive GCs being formed in larger star bursts at larger redshifts \citep{Bastian2008}. 
A significant fraction of these GCs will survive through the following evolution of the 
galaxy \citep{Fall2001}. A detailed investigation of the behavior of \sn\ 
and its dependence on the environment and formation history of the host galaxy is beyond the 
scope of this paper. However, presence of the \sn-age correlation (at least in our sample) 
significantly complicates the separation of the effects of these two factors on the LMXB population.

\subsection{Conclusion}

There are a number of uncertainties associated with measurements of stellar age and \sn. Many of these, for example, age gradients, can be alleviated in principle with more sophisticated analysis of the optical data, e.g., with spatially resolved stellar age maps. On the other hand, proper use of the improved optical data would require better statistics in the X-ray data. Indeed, as one can see from Fig. \ref{fig:n_age_sn}, the vertical error bars, which are determined by the accuracy of the X-ray data, are often nearly as big as the uncertainties in the stellar ages. This requires dedicated effort in collecting optical and X-ray data of adequate quality for sufficiently large numbers of galaxies to allow statistically meaningful analysis. Further progress may be possible by comparing the spatial distributions of X-ray sources with the SAURON stellar age maps \citep[cf.][]{Sht2007}. Other deficiencies, for example, the failure of the SSP assumption in regions with complex age content, require different data analysis approaches to be considered. 

In the present paper, we used the currently available data and a somewhat simplistic approach, in the spirit of the SSP assumption, to investigate for the first time the sign and the amplitude of the secular evolution of LMXB populations with cosmic time. 
Obviously, the age and \sn\ uncertainties affect the appearance of the data points in the age--\fxlf\ and \sn --\fxlf\ planes and contribute to their dispersion. However, the fact that we detect statistically significant correlations between parameters of interest suggests that the uncertainties do not completely wash out the effects of the age and GC content on the population of LMXBs in galaxies. This is further supported by the correlation between \sn\ and stellar age, as the errors in these quantities are uncorrelated.

\section{Discussion}
\label{sec:discussion}
\label{sec:implication}

\subsection{GCs vs stellar age}

The dependence of the LMXB population on stellar age and GC content cannot be easily separated due to the existence of a strong correlation between these two parameters in our sample.
Results of the linear regression analysis suggest that about $\sim 2/3$ of the amplitude of the \fxlf\ variation is due to stellar age dependence and $\sim 1/3$ is caused by changing GC content. However, because of the large scatter of the points and corresponding very large value of the $\chi^2$, this result should be interpreted  with caution and any firm conclusion would be premature. 

The X-ray binaries dynamically formed in GCs can reside in their parent clusters or be expelled into the field, either due to kicks and/or tidal interactions or during the destruction of the parent cluster. The former could in principle be identified and removed from the sample, provided that clean and complete GC lists are available. This, however, is not possible due to 
limitations of the GC data. A significant part of which was obtained by the ground-based 
facilities and suffers from incompleteness and foreground star contamination. Although HST data is available for a number of galaxies, it also has limited completeness. Besides, in the luminosity range of interest, X-ray sources in GCs do not account for  more than $\sim 30-40\%$ of the LMXB population \citep{Zhang2011},  which is hardly sufficient to explain the amplitude of the observed dependence 
(right-hand panel in Fig.\ref{fig:n_age_sn}). The contribution of binaries that are dynamically formed 
in GCs and expelled into the field is unclear. Based on the correlations of the specific frequency  of LMXBs with that of GCs, it has been argued that the entire population of LMXBs has been  formed dynamically in GCs \citep[e.g.,][]{White2002}. Our results, however, suggest that an important 
factor is the evolution of the LMXB population with time. Part of the correlation of the specific frequency 
of LMXBs with that of GCs is real, due to the contribution of sources formed dynamically 
(and mostly retained) in GC. Another part is, however, an artifact of the dependence of the 
GC content on the age of the galaxy.  

The importance of age dependence is further supported by the comparison of the XLF of young and old 
galaxies (Fig.\ref{fig:xlfage}). \citet{Zhang2011} have shown that the XLF of GC sources in the 
$\log(L_X)\ga 37$ luminosity range is flatter than that of the field sources (see, e.g., their Fig.7). 
Therefore, as older galaxies have larger GC content, i.e., a larger fraction of dynamically 
formed binaries, their XLF should be expected to be flatter than the XLF of the young galaxy sample, if the main increase is due to the dynamically formed sources. This 
prediction is at odds with our observations: the XLF of old galaxies is steeper than that of 
the young galaxies (see also \citet{Kim2010}). This suggests  that the effects of the evolution of the LMXB population are at least comparable to the effect of the increased GC content, if not stronger.

\subsection{Comparison with Fragos et al. calculations}
\label{sec:fragos}

Our results appear to be at odds with the population synthesis calculations of \citet{Fragos2008}, who predicted a significant  decrease of the LMXB population  with time. In particular, they predict more than an order of magnitude decrease of the number of LMXBs in the age interval from 5.5 Gyrs to 9.5 Gyrs. Their prediction includes the $\log(L_X)\ga 37.5$ luminosity range, well covered by the  Chandra data used here, so that such a significant change in the specific number of sources would clearly reveal itself in our analysis. This is illustrated, for example, by the  comparison of our Fig. \ref{fig:xlfage} with the Fig.3 in \citet{Fragos2008}. Along the same lines, in more recent calculations \citet{Fragos2012} considered the overall evolution of the X-ray binary populations with cosmic time and came to a similar conclusion: that the specific luminosity of LMXBs per unit stellar mass in the Universe is decreasing by about an order of magnitude between the redshift $z=1$ and $z=0$. 

The reason for this discrepancy is not clear. On the one hand, the Fragos et al. calculations consider primordial binaries only and do not include dynamically formed systems in GCs. Therefore, their results could be reconciled with our observations if one assumed that the absolute majority of LMXBs are of dynamical origin. Quantitatively, in order to allow more than a ten-fold decrease in the numbers of primordial LMXBs between 5.5 and 9.5 Gyrs and still be compatible with our observations,  the contribution of primordial systems should be less than a few per cent in the $\sim 10$ Gyrs old galaxies. This would also imply that, on average, dynamically formed LMXBs contribute $\ga 95-99\%$ to the LMXB population at the redshift $z=0$. We consider this possibility unlikely because it would contradict the evidence presented in this paper and in other studies \citep[e.g.,][]{Juett2005,Kundu2007,Voss2009,Zhang2011}.

Another important caveat is that \citet{Fragos2008} considered coeval stellar populations. Similar  considerations complicate stellar age determinations, so the discussion of Sect.~\ref{sec:caveats_age} applies here fully. 
Quantitatively, however, significant deviations from the SSP assumption in galaxies with young SSP ages would be required in order to account for the observed discrepancy between theory and our observations. Indeed, the single-starburst calculations by \citet[][in Fig. 2]{Fragos2012}  predict a $\approx 100$-fold decrease in the specific LMXB luminosity between 2 and 10 Gyrs. Such a drop in the LMXB populations can be reconciled with our results only by assuming that the fraction of young stars in galaxies with the youngest SSP ages (e.g., NGC 4382 -- 1.6 Gyrs and NGC 5866 -- 1.8 Gyrs) is actually rather small, $\la 10^{-2}$. Detailed data about the star-formation history and stellar age structure of galaxies with young SSP ages is needed in order to see if this possibility is feasible.

\section{Summary}
\label{sec:summary}

The main goal of this paper was to study the dependence of the population of LMXBs on  stellar age. To this end, we collected 20 nearby early-type galaxies, which were observed by \chandra\ to sufficient depth, and had the stellar age measured. 

\begin{enumerate}
\item

We found that older galaxies tend to host more LMXBs per unit stellar mass than younger ones (Fig.\ref{fig:n_age_sn}). The correlation has large scatter, with the points occupying the dynamical range by 
a factor of 4. When averaged over young ($t<6$ Gyrs) and old ($t>6$ Gyrs) subsamples, the specific frequency 
of LMXBs with $L_X>5\cdot10^{37}$ erg/s varies from  $4.17\pm0.27$ to $6.27\pm0.26$ per $10^{10}~M_\odot$. 
Interpretation of this dependence is complicated by the rather strong correlation between the GC 
content of the galaxy and its stellar age. We presented evidence suggesting that an important factor is the 
intrinsic evolution of the populations of LMXBs with time. Its effect is enhanced by the larger GC content of older galaxies, resulting in larger numbers of dynamically formed binaries in them.

\item
Our results appear to challenge recent population synthesis calculations by \citet{Fragos2008, Fragos2012}, predicting a more than $\sim$ten-fold decrease of the primordial LMXB population between $\sim 5.5$ and 9.5 Gyrs (a hundred-fold decrease between 2 and 10 Gyrs). This discrepancy  can be understood  under rather extreme assumptions about the contribution of the dynamically formed LMXBs and/or the fraction of truly young stellar populations in the galaxies with young SSP ages. The caveats to the comparison of our data with calculations of  \citet{Fragos2008, Fragos2012}  are discussed in Sect.~\ref{sec:fragos}.

\item
There is clear evolution of the XLF with age: the one of older galaxies is steeper 
in the entire studied luminosity range, $\log(L_X)\ga 37.5$, than the one 
of younger galaxies (Fig.~\ref{fig:xlfage}).  A similar result was also reported by \citet{Kim2010}.  

\item
Young galaxies host a significant population of (ultra-) luminous X-ray sources with luminosity exceeding 
$10^{39}$ erg/s. We estimate their specific frequency of  $\approx 8.8\pm3.2$ sources  per $10^{12}~M_\odot$ 
in the young subsample. Such sources are significantly less frequent in the old subsample ($\approx 1$ 
source against $\approx 7$), with the 90\% upper limit  of $\approx 2.9$ sources per $10^{12}~M_\odot$ 

\item 
As a byproduct of this study, we compiled a list of six black hole candidates in GCs, of which 
 five were previously known and one is identified for the first time (Table~\ref{tab:ulx}).

\end{enumerate}

\begin{acknowledgements} 
This research made use of \chandra\ archival data provided by
\chandra\ X-ray Center and the 2MASS Large Galaxy Atlas data provided by
NASA/IPAC infrared science archive. {\'A}kos Bogd{\'a}n acknowledges 
support provided by NASA through Einstein Postdoctoral Fellowship 
grant number PF1-120081 awarded by the Chandra X-ray Center, which 
is operated by the Smithsonian Astrophysical Observatory for NASA
under contract NAS8-03060. We also wish to thank Rasmus Voss, 
Andrew Cooper, Diederik Kruijssen, Junhua Gu, and Jingying Wang 
for the discussions which have greatly improved the quality of 
this paper.  

\end{acknowledgements}

\bibliographystyle{aa}
\bibliography{ms}

\end{document}